\documentclass[]{spie}  
\usepackage{soul}
 
\usepackage{amsmath,amsfonts,amssymb}
\usepackage{graphicx}
\usepackage[colorlinks=true, allcolors=blue]{hyperref}
\usepackage[table,xcdraw]{xcolor}
\usepackage{gensymb}

\title{Simulations of polarimetric observations of debris disks through the Roman Coronagraph Instrument}

\author[a]{Ramya M Anche}
\author[a]{Ewan S. Douglas}
\author[a,b]{Kian Milani}
\author[a,b]{Jaren Ashcraft}
\author[c]{John Debes}

\affil[a]{Steward Observatory, University of Arizona, 933N Cherry Avenue, Tucson, Arizona, 85721, USA}
\affil[b]{James C. Wyant College of Optical Sciences, University of Arizona, 933N Cherry Avenue, Tucson, Arizona, 85721, USA}
\affil[c]{Space Telescope Science Institute, 3700 San Martin Drive,
Baltimore, MD 21218, USA}

\authorinfo{Further author information: (Send correspondence to R.M.A)\\R.M.A: E-mail: ramyaanche@arizona.edu}

\begin{document} 
\maketitle

\begin{abstract}
The Roman coronagraph instrument will demonstrate high-contrast imaging technology, enabling the imaging of faint debris disks, the discovery of inner dust belts, and planets. Polarization studies of debris disks provide information on dust grains' size, shape, and distribution. The Roman coronagraph uses a  polarization module comprising two Wollaston prism assemblies to produce four orthogonally polarized images ($I_{0}$, $I_{90}$, $I_{45}$, and $I_{135}$), each measuring 3.2 arcsecs in diameter and separated by 7.5 arcsecs in the sky. The expected RMS error in the linear polarization fraction measurement is 1.66\% per resolution element of 3 by 3 pixels. We present a mathematical model to simulate the polarized intensity images through the Roman CGI, including the instrumental polarization and other uncertainties. We use disk modeling software, MCFOST, to model $q$, $u$, and polarization intensity of the debris disk, Epsilon-Eridani. The polarization intensities are convolved with the coronagraph throughput incorporating the PSF morphology. We include model uncertainties, detector noise, speckle noise, and jitter. The final polarization fraction of 0.4$\pm$0.0251 is obtained after the post-processing. 
\end{abstract}

\keywords{Roman CGI, Debris disks, Polarization observations, coronagraphs, Polarimetric calibration}

\section{INTRODUCTION}
Debris disks are predominantly made of dust and planetesimals and are found to be an integral part of the evolution of the planetary systems\cite{hughes2018debris,backman2004debris}. The Asteroid belt and the Kuiper belt form the debris disk of our solar system along with the zodiacal dust \cite{wyatt2016insights}. Debris disks may be polarized due to Rayleigh/Mie scattering by dust particles or asymmetry or the dichroic extinction of the dust grains in the presence of magnetic fields. Polarization observations of debris disks enable us to constrain their structure and properties of the dust grains such as composition, size, shape, and distribution\cite{crotts2021deep,hull2022polarization,chen2020multiband,engler2017hip}.
Polarization observation of several debris disks has been carried out using the Gemini Planet Imager (GPI)\cite{macintosh2014first} at the Gemini Telescope, and  SPHERE-ZIMPOL \cite{beuzit2006sphere} at the Very Large Telescope (VLT) in the near IR wavelengths. 
Hubble Space Telescope Advanced Camera for Surveys coronograph has been used to observe the polarization of a few debris disks (HD61005 \cite{maness2009hubble}, and AU Mic \cite{graham2007signature}) in the Johnson-Cousins V band filter. The multi-band polarization observations will allow us to constrain better the dust properties and disk structure of the debris disks. 
\par
The upcoming Nancy Grace Roman Space Telescope Coronograph instrument (CGI) will facilitate the polarization observations of fainter debris disks (reaching contrast level of $\sim10^{-8}$) around the nearby stars in addition to the high-contrast and high-resolution imaging of exoplanets. The polarimetric module consists of two Wollaston prisms, each producing two orthogonally polarized images, separated by 7.5'' on the sky. The imaging polarimetry is available for the narrow field (525nm), and wide-field (825nm) modes \cite{mennesson2021roman}. In this paper, we describe a mathematical model for simulating the polarization observations of debris disk through the Roman CGI. In our simulations, we have considered the debris disk around a nearby (early) sun-like star, Epsilon Eridani ($\epsilon$ Eridani). 

The next section provides an overview of the mathematical model and a snapshot of each stage's results. The following sections describe each step in the mathematical model with the corresponding outputs. Finally, the conclusions and future work are presented in the discussion section.
\label{sec:intro}  
\section{Mathematical model}
\begin{figure}[!h]
\begin{center}
\fbox{\includegraphics[width=0.95\linewidth]{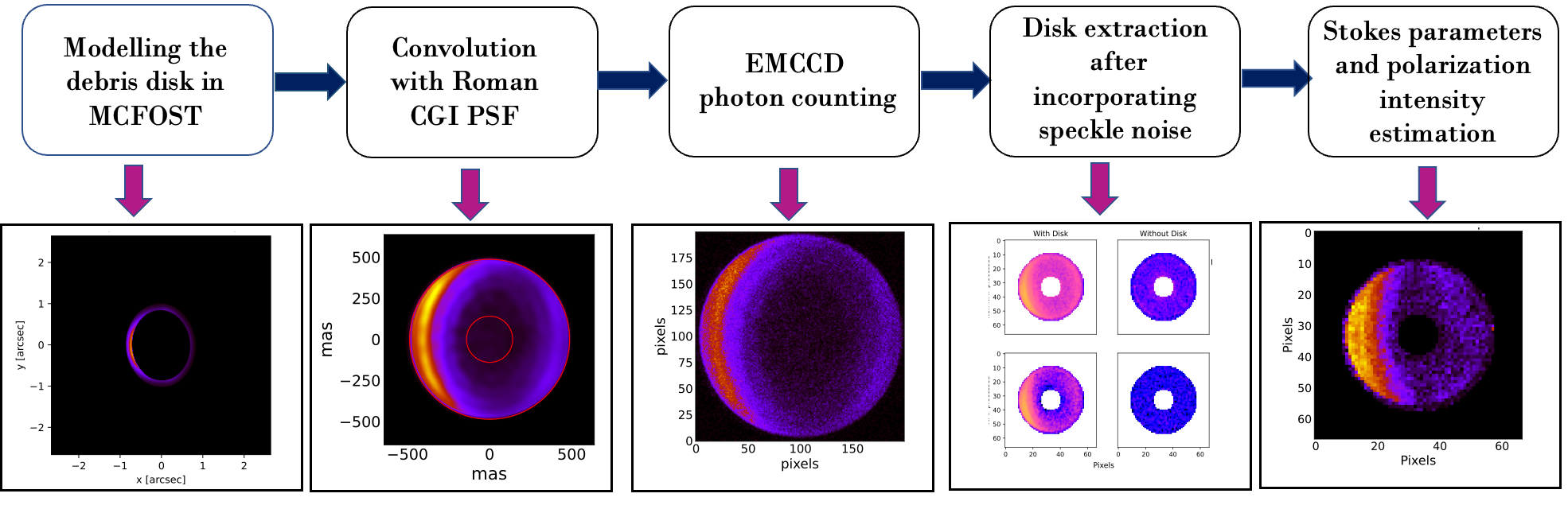}}
\caption{Mathematical model for simulating the polarization of debris disk through Roman CGI}
\label{model}
\end{center}
\end{figure}

In this section, we describe the overview of the mathematical model developed for the simulation of polarization observation of debris disk Epsilon-Eridani as shown in Figure \ref{model}. First, the disk is modeled using \href{https://ipag.osug.fr/~pintec/mcfost/docs/html/overview.html}{MCFOST} to obtain the scattered light intensity and Stokes parameters. Next, the disk intensities are convolved with the Roman CGI PSF obtained using the \href{http://proper-library.sourceforge.net/}{PROPER} models. Further, the EMCCD photon counted images are generated, including the EMCCD gain and noise characteristics. Next, we add the speckle and jitter noise using \href{https://roman.ipac.caltech.edu/sims/Coronagraph_public_images.html#CGI_OS9}{Os9} simulations of Roman CGI and perform disk extraction. The final step is the estimation of Stokes parameters and corresponding errors after multiplying the instrument Mueller matrix. Each stage of this model is described in the following sections.

\section{Modelling of Epsilon-Eridani in MCFOST}
Epsilon Eridani is a star similar to the early sun; at a distance of 3.2pc with T=5100K, Mass=0.82\(M_\odot\) and Radius= 0.88\(R_\odot\). The debris disk around $\epsilon$ Eridani has been observed at multiple wavelengths, yet, the disk parameters and details of the structure are debatable. The debris disk has been detected as an unresolved excess around the star with two or three components: a warm inner disk with  $\sim$ two narrow belts according to $Su~et.~al (2017)$\cite{su2017inner} (that has not yet been resolved) and a cold outer disk imaged with both ALMA\cite{booth2017northern} and other Sub-millimeter instruments \cite{backman2008epsilon,greaves2005structure}.

\begin{figure}[!h]
\begin{center}
\fbox{\includegraphics[width=0.7\linewidth]{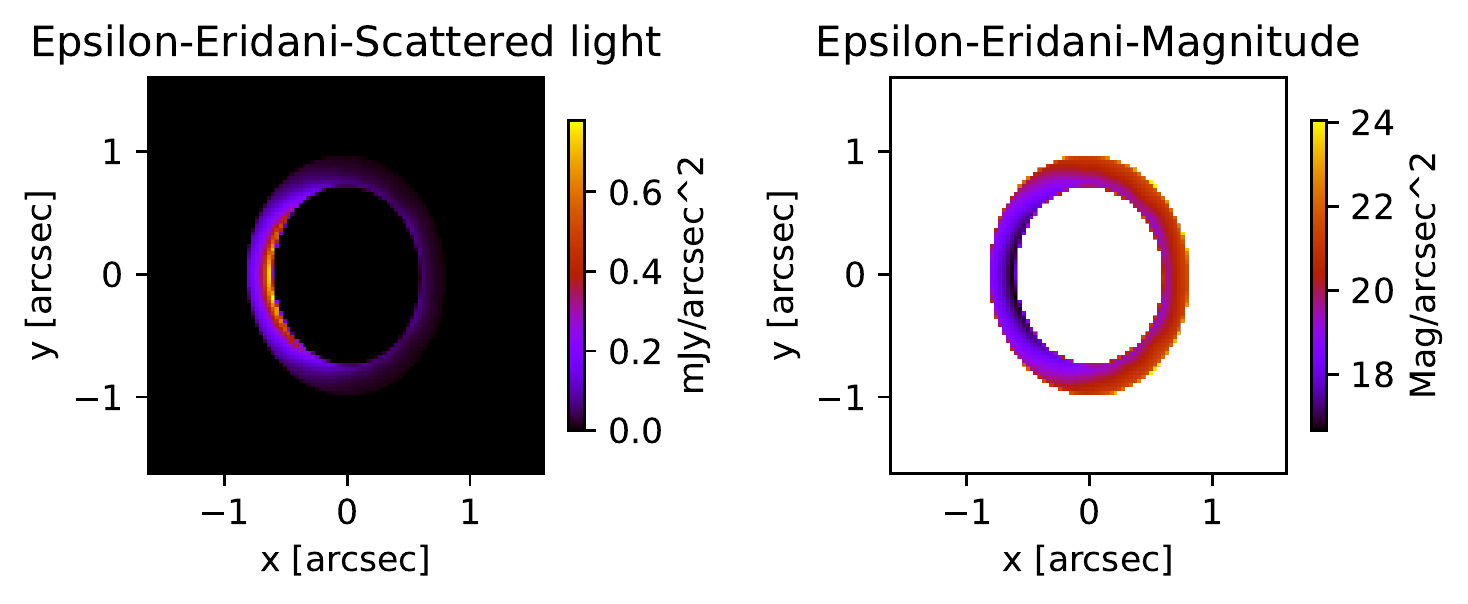}}
\caption{Scattered light brightness and magnitude obtained (for the innermost disk) using \href{https://ipag.osug.fr/~pintec/mcfost/docs/html/overview.html}{MCFOST} at $\lambda$=575nm }
\label{scattered light}
\end{center}
\end{figure}

\begin{figure}[!h]
\begin{center}
\fbox{\includegraphics[width=0.7\linewidth]{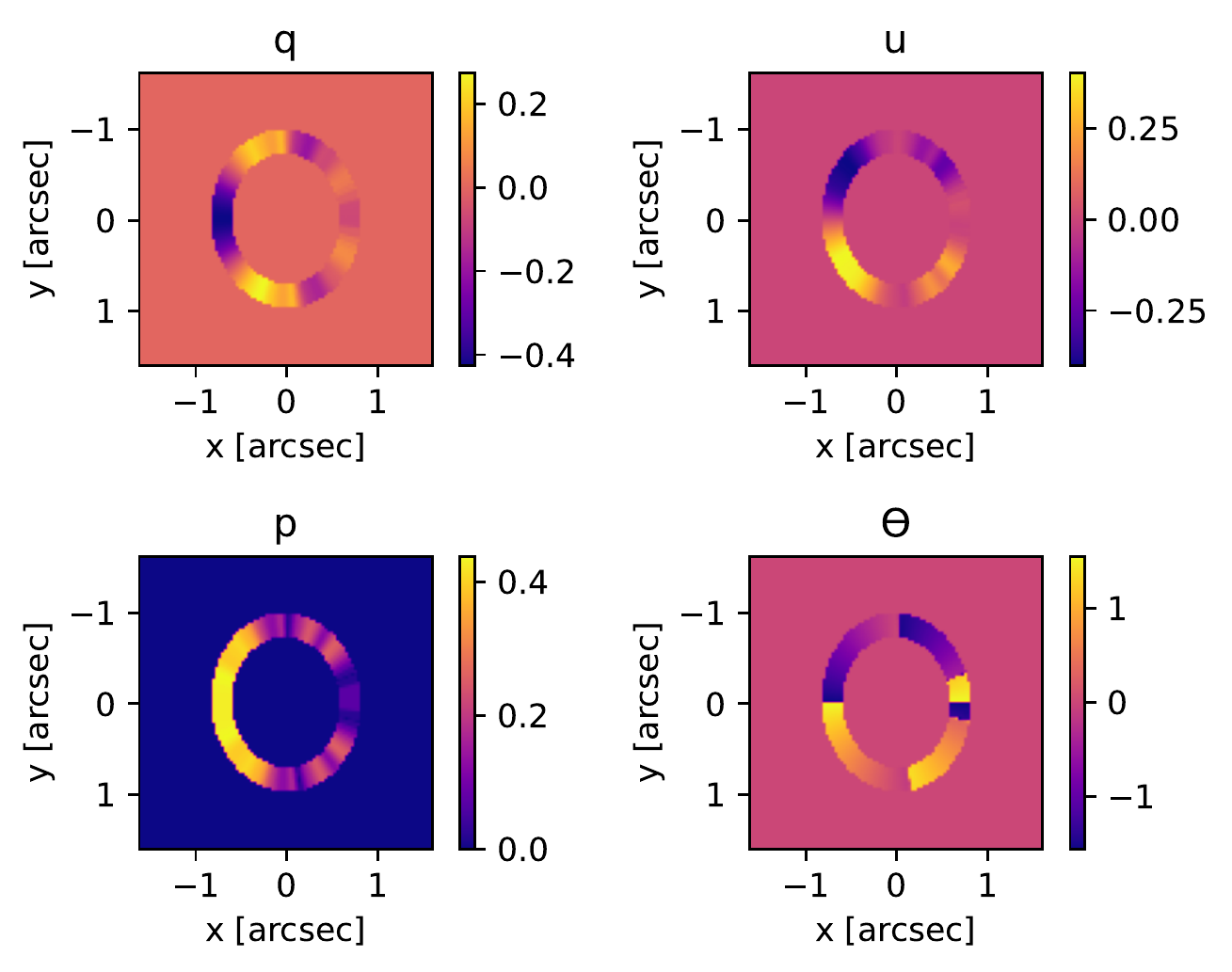}}
\caption{Stokes parameters obtained (for the innermost disk) using  \href{https://ipag.osug.fr/~pintec/mcfost/docs/html/overview.html}{MCFOST} at $\lambda$=575nm}
\label{stokes}
\end{center}
\end{figure}

The Roman CGI is expected to resolve the innermost disk around $\epsilon$ Eridani and provide clarity regarding the size and location of the inner disk. We have used the disk parameters and dust composition from $Su~et.al(2017)$\cite{su2017inner} to model the inner disk of the $\epsilon$ Eridani. The parameters provided in Table 1 are used in  \href{https://ipag.osug.fr/~pintec/mcfost/docs/html/overview.html}{MCFOST} to obtain the scattered light and Stokes parameters (for linear polarization: $q$, and $u$) images at a wavelength of 575nm and inclination of 34\textdegree. We used a 20.8 mas/pixel scale with 256$\times$256 pixels and the Mie scattering model to account for the polarization. The scattered light images and Stokes parameters are shown in Figures \ref{scattered light} and \ref{stokes} respectively. 

\begin{table}[!h]
\begin{center}
\begin{tabular}{ccc}
\hline
\multicolumn{1}{c}{Parameters} & \multicolumn{2}{c}{Eps-eri   model} \\ \hline
Disks                          & \multicolumn{2}{c}{2 disk}          \\ \hline 
Disk extent (AU)               & 1.5-2            & 8-20$^\ast$           \\
Scale Height                   & 0                & 0                \\
Dust Mass (Mo)                 & 1.35*$10^{-12}$       & 1.00*$10^{-12}$       \\
Surface density Exponent       & -10 to -10       & -10 to -10       \\
Grain sizes(microns)           & 1-1000 $e^{3.65}$  & 1-1000 $e^{3.65}$  \\
Grain composition              & Astrosilicates   & Ice-silicates  
\\ \hline
\end{tabular}
\label{mcfost-para}
\end{center}
\caption{Parameters used in the MCFOST modelling of $\epsilon$ Eridani. $\ast$ is not visible in figure \ref{scattered light}}
\end{table}

\begin{figure}[!h]
\begin{center}
\fbox{\includegraphics[width=0.9\linewidth]{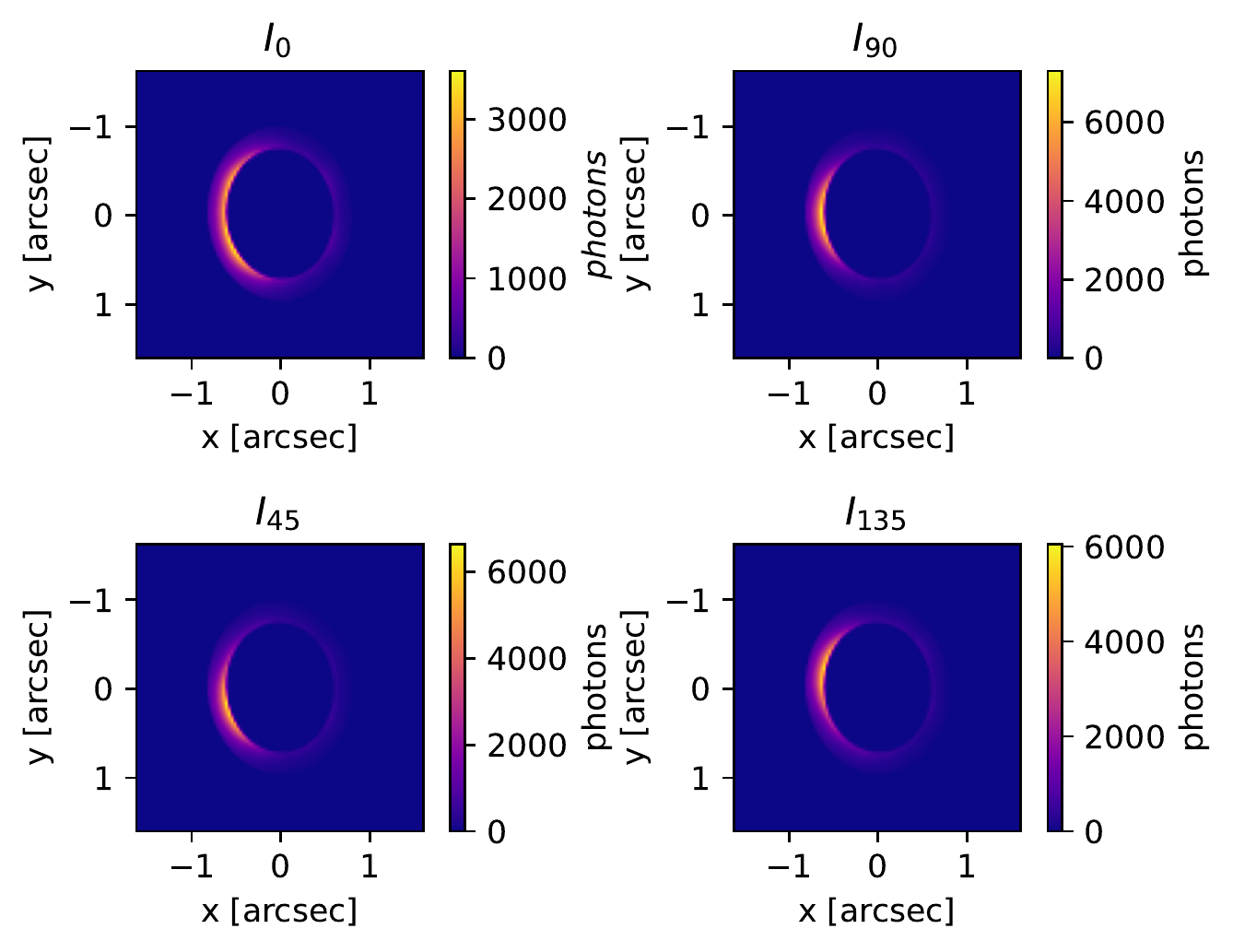}}
\caption{Polarization intensities estimated using $p$, $\theta$ and Scattered light brightness at $\lambda$=575nm}
\label{polint}
\end{center}
\end{figure}

The scattered light brightness and $p$ and $\theta$ are used to obtain the four intensities ($I_0$, $I_{90}$,$I_{45}$ and $I_{135}$) as below:
\begin{align}
I_0=Nsrct*(1+p*\cos(2*\theta))/2 \quad
I_{90}=Nsrct*(1-p*\cos(2*\theta))/2 \\
I_{45}=Nsrct*(1+p*\sin(2*\theta))/2 \quad
I_{135}=Nsrct*(1-p*\sin(2*\theta))/2
\end{align}
where $Nsrct$ correspond to the total number of photons.The four intensities are shown in Figure \ref{polint}. These intensities are then convolved with the Roman CGI PSFs as explained in the next section.

\section{Generating Roman PSFs}

Point Response Functions (PRFs) are generated for the Hybrid Lyot Coronograph mode at 575nm with a field of view of 0.14''-0.45'' using \href{http://proper-library.sourceforge.net/}{PROPER} model of the Roman CGI. We set the polarization axis as the mean of all polarization modes in the \href{http://proper-library.sourceforge.net/}{PROPER} model (polaxis=10). 
Convolution is done via a matrix-vector multiplication in a process utilizing the PRFs of field angles for each pixel in the disk model array
\cite{milani2020fastersims}.

\begin{figure}[!h]
\begin{center}
\fbox{\includegraphics[width=0.28\linewidth]{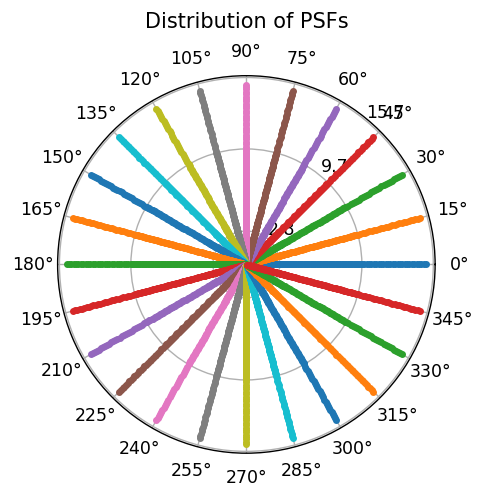}}
\fbox{\includegraphics[width=0.308\linewidth]{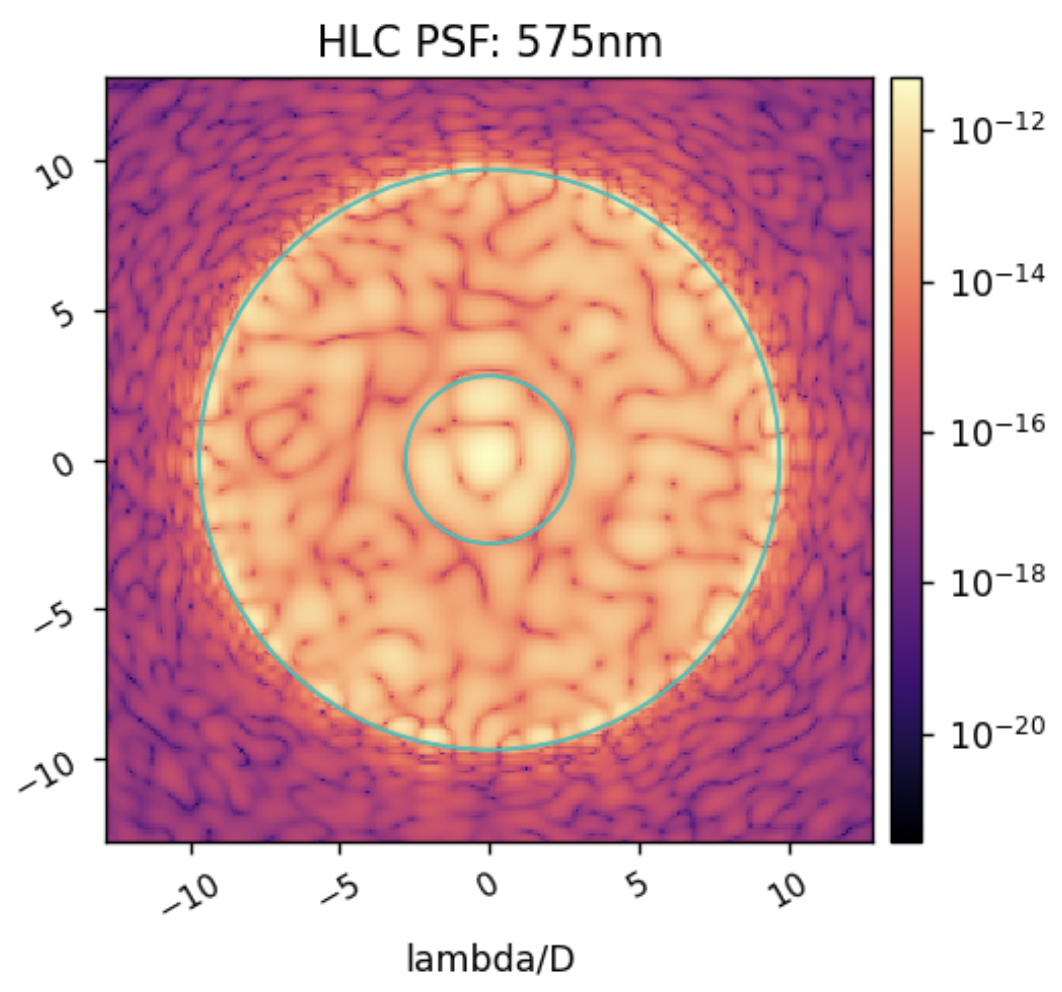}}
\caption{Distribution of PRFs used to create image simulations (left). On-axis monochromatic PSF for the HLC (right).}
\label{prfs}
\end{center}
\end{figure}
\begin{figure}[!ht]
\begin{center}
\fbox{\includegraphics[width=0.9\linewidth]{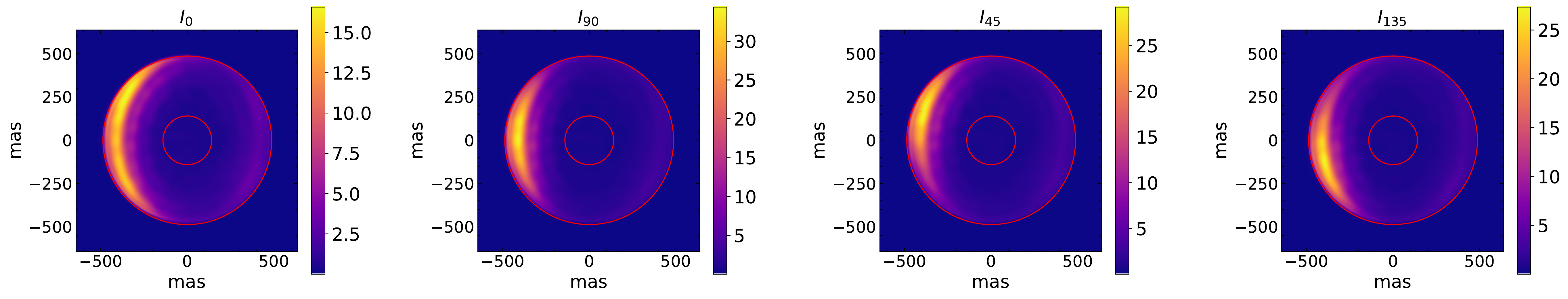}}
\caption{Polarization intensities (photons) after convolving with the PSF}
\label{polintsim}
\end{center}
\end{figure}

In addition, PRFs for field angles in both image dimensions are utilized to account for the roll angle. This is illustrated in Figure \ref{prfs} with the distribution of PRFs used for the HLC. Finally, PRFs beyond the OWA are included to incorporate the scattering of light from regions of the disk outside the OWA. The disk images of $I_0$, $I_{90}$, $I_{45}$, and $I_{135}$ are shown in Figure \ref{polintsim} after convolving with the Roman PSFs.
\section{EMCCD simulation}
\begin{figure}[!h]
\begin{center}
\includegraphics[width=0.9\linewidth]{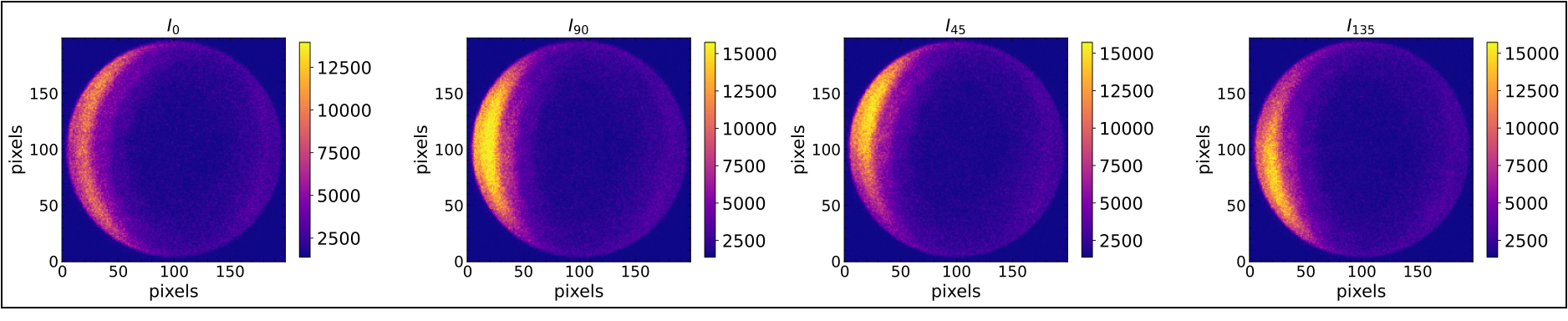}
\caption{Polarization intensities for one of the EMCCD frames from the stack}
\label{emccdframe}
\end{center}
\end{figure}

The Roman CGI will use \href{https://www.teledyneimaging.com/en/aerospace-and-defense/products/sensors-overview/ccd/ccd201-20/}{e2v CCD201-20} which is a back-illuminated electron-multiplying CCD sensor consisting of 1024$\times$ 1024 pixels of 13$\mu$m in size. It can be operated in low gain ($<$1000) and high gain ($>$1000) mode.
To convert the disk polarization intensities to photo electrons \href{https://github.com/wfirst-cgi/emccd_detect}{EMCCD detect} \cite{nemati2020photon} is used.  We simulate a stack of 50 EMCCD frames for each of the intensity component with an exposure time=5s/frame, gain=800 incorporating bias=10000$e^-$, dark current =0.0028, $e^-/pix/s$, read noise=100 $e^-$. Figure \ref{emccdframe} shows one of the frames at the EMCCD for all the four polarization intensities.    

\section{Disk extraction and estimating Stokes parameters}
\begin{figure}[!h]
\begin{center}
\includegraphics[width=0.6\linewidth]{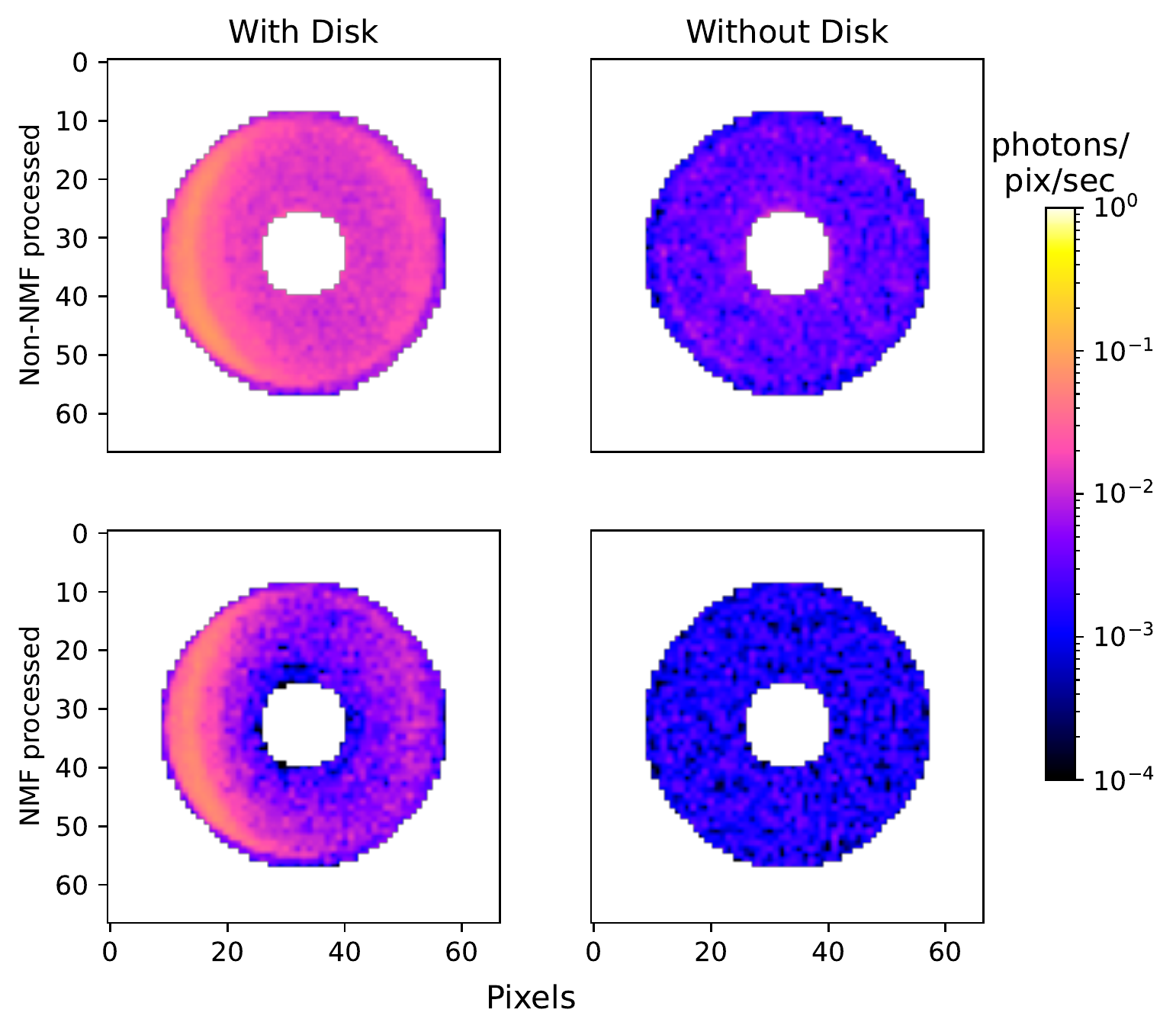}
\caption{Top: raw, unprocessed coronographic images with and without disk injection, Botton: NMF speckle subtraction post-processing applied - for $I_0$ component}
\label{extracteddisk}
\end{center}
\end{figure}

The integrated modeling team at JPL creates Observing Scenario simulations to generate the simulated science data for the HLC and SPC coronographs using the most recent version of the observation strategy.
\href{https://roman.ipac.caltech.edu/sims/Coronagraph_public_images.html#CGI_OS9}{OS9} is the ninth public release of Roman post-coronagraph simulated science images, which includes end-to-end Structural Thermal Optical Performance (STOP) model of the Roman observatory, coronagraph masks, diffraction, wavefront control, and detector noise and also jitter. We perform the disk processing, and extraction technique as explained in $Douglas~et.~al (2022)$ \cite{douglas2022sensitivity}. The photon-counting procedure is operated on all the intensity images with a threshold of 1.1*readnoise, following the steps from $Nemati(2020)$ \cite{nemati2020photon}. We include optical model uncertainty factors (MUFs) in our analysis obtained from the OS9 repository. Although we use the Non-negative matrix factorization (NMF) method \cite{ren2018non} to subtract the PSF components while performing disk extraction, we can obtain the required SNR without using the NMF as well. Figure \ref{extracteddisk} shows the unprocessed coronographic images with and without disk and post-processed disk with NMF speckle subtraction for one of the polarization intensity components.
Once the photon counted images are obtained for all the four intensity components, the Stokes parameters and corresponding errors are estimated as follows \cite{ramaprakash2019robopol}:
\begin{align}
q'{}=\frac{I_{0^\circ}-I_{90^\circ}}{I_{0^\circ}+I_{90^\circ}}  \hspace{0.2cm} \quad  
u'=\frac{I_{45^\circ}-I_{135^\circ}}{I_{45^\circ}+I_{135^\circ}}
\end{align}
\begin{align}
\sigma_{q}=\sqrt{\frac{4\left(I_{90^\circ}^{2} \sigma_{I{0^\circ}}^{2}+I_{0^\circ}^{2} \sigma_{I_{90^\circ}}^{2}\right)}{\left(I_{0^\circ}+I_{90^\circ}\right)^{4}}}, \hspace{0.2cm} \quad
\sigma_{u}=\sqrt{\frac{4\left(I_{45^\circ}^{2} \sigma_{I_{135^\circ}}^{2}+I_{135^\circ}^{2} \sigma_{I_{45^\circ}}^{2}\right)}{\left(I_{45^\circ}+I_{135^\circ}\right)^{4}}}
\end{align} 
where, $I_{0^\circ}, I_{90^\circ}, I_{45^\circ}, I_{135^\circ}$ are the four intensities in polarized beams and $\sigma_{I_{0^\circ}}, \sigma_{I_{90^\circ}}, \sigma_{I_{45^\circ}}, \sigma_{I_{135^\circ}}$ are corresponding uncertainties. 
Although all the instrument effects are considered in the PSF convolution and OS9 simulations, we have not incorporated the polarization changes introduced by the Roman telescope and instrument optics. Any optical component in the system can either introduce or modify the incoming source polarization causing instrumental polarization, or polarization crosstalk \cite{tinbergen2005astronomical,anche2018analysis}. These effects can be accounted for using the 4$\times$4 Mueller matrix of the system, as explained in the next section. 

\section{Mueller matrix of the CGI instrument and estimating final stokes parameters}
\begin{figure}[!h]
\begin{center}
\fbox{\includegraphics[width=0.85\linewidth]{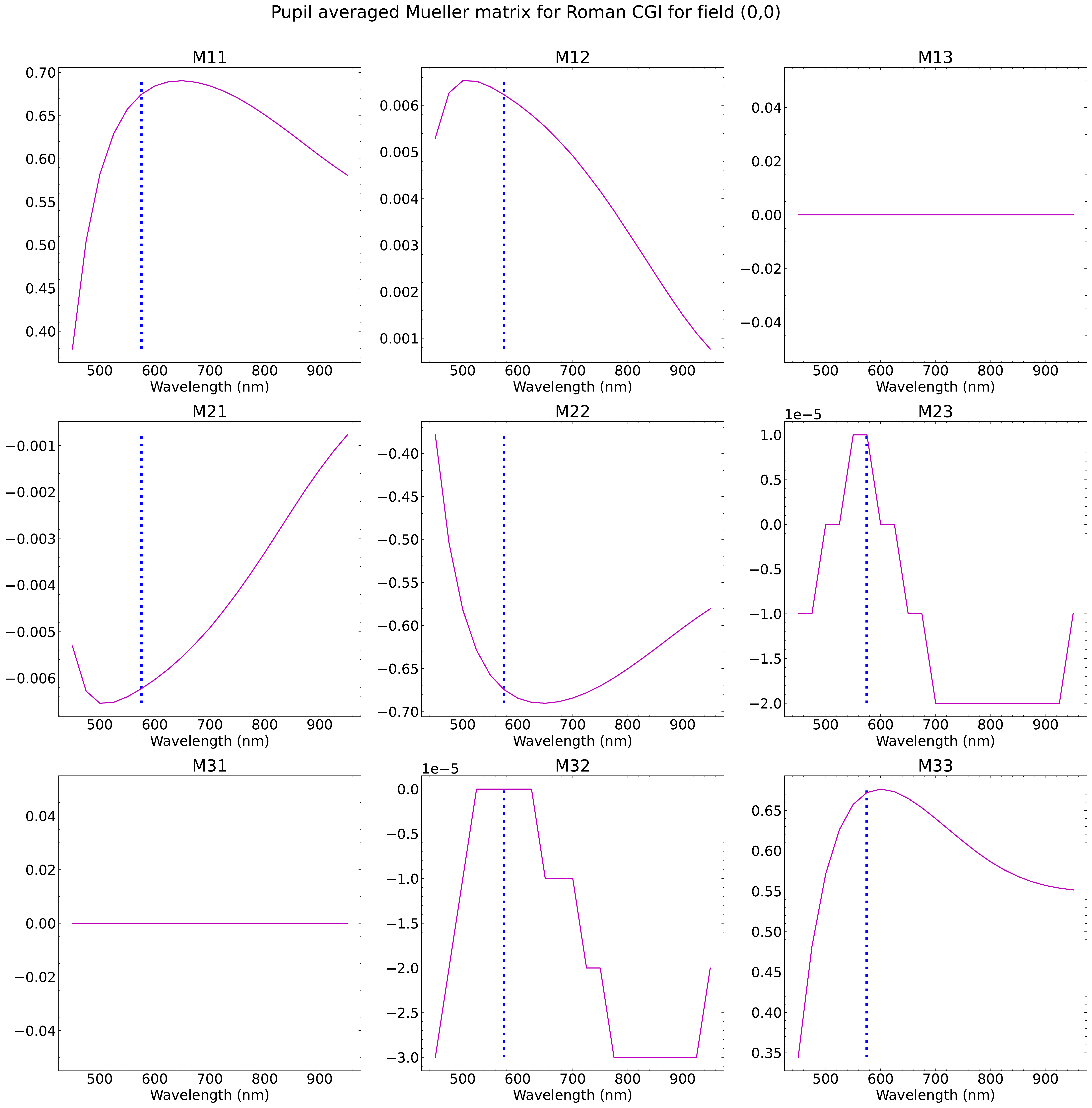}}
\caption{Pupil averaged Mueller matrix for Roman CGI for on-axis field obtained from  \href{https://roman.ipac.caltech.edu/sims/Coronagraph_inst_param_data_more.html}{IPAC} The dotted blue line shows the values of the matrix elements used in this analysis at $\lambda$=575nm}
\label{Muellermatrix}
\end{center}
\end{figure}

\begin{figure}[!h]
\begin{center}
\fbox{\includegraphics[width=0.63\linewidth]{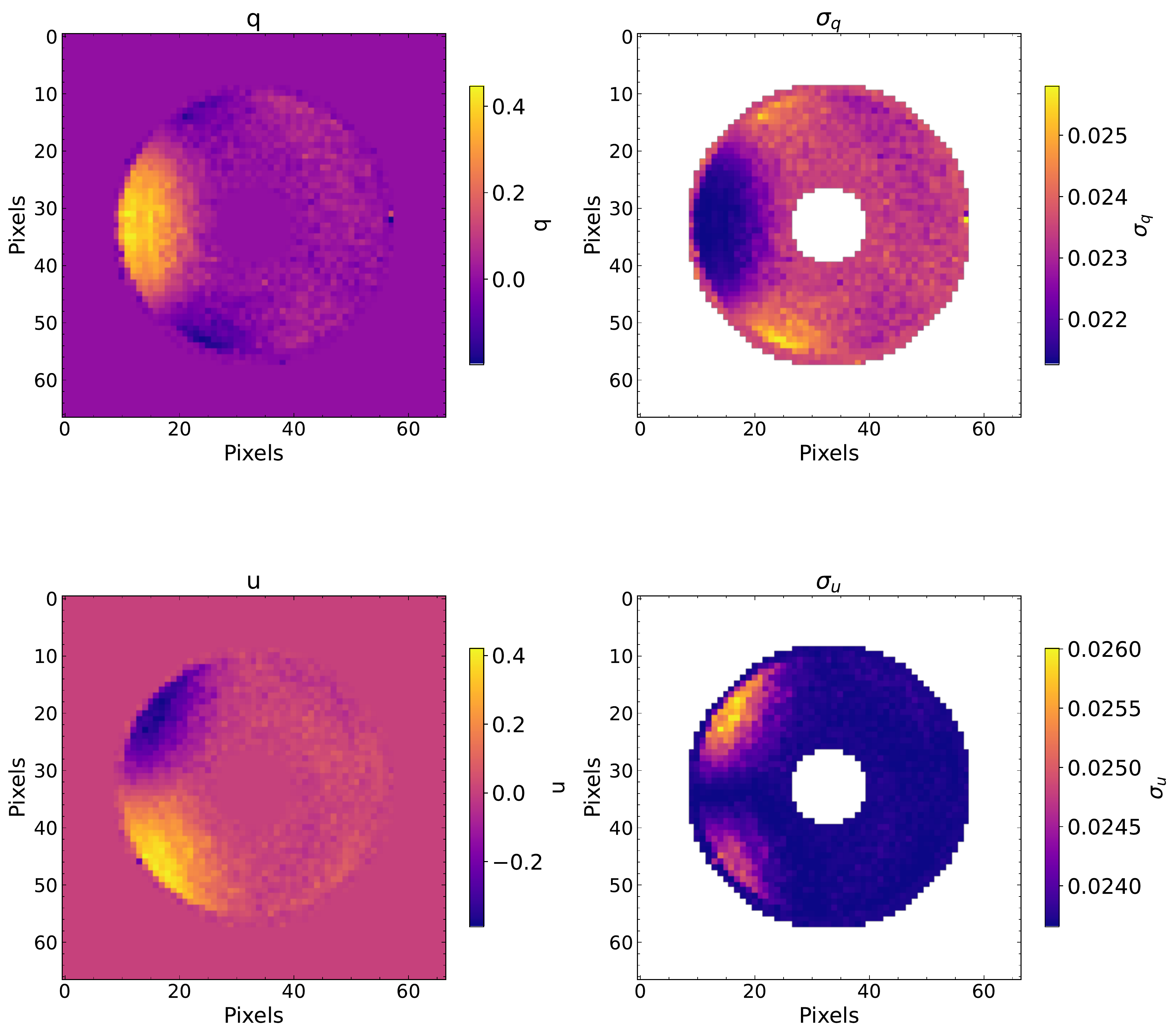}}
\caption{Stokes $q$ and $u$ and their corresponding errors estimated from the extracted disks}
\label{stokesqu}
\end{center}
\end{figure}

\begin{figure}[!h]
\begin{center}
\fbox{\includegraphics[width=0.63\linewidth]{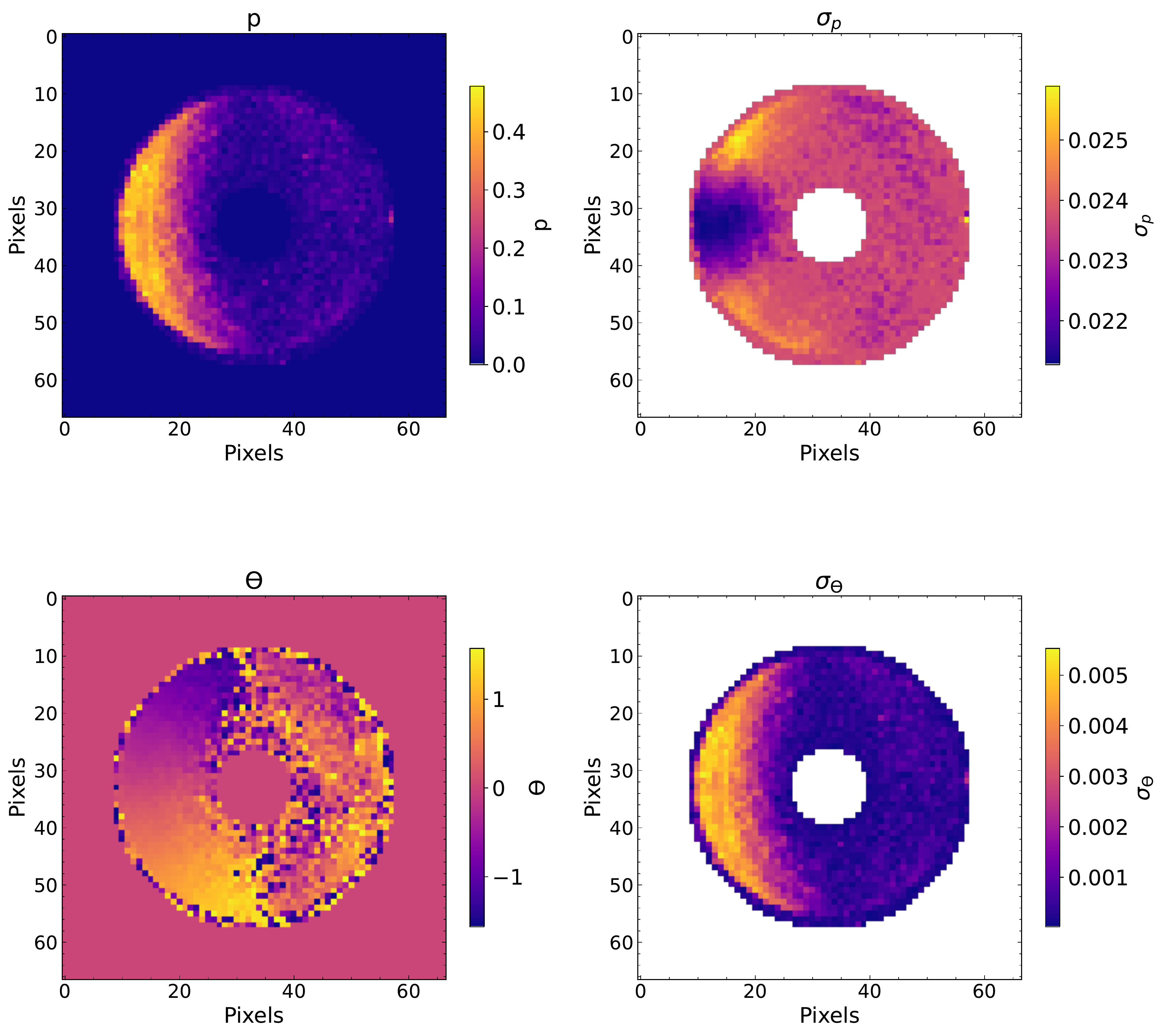}}
\caption{Stokes $p$ and $\theta$ and their corresponding errors estimated from the extracted disks}
\label{ptheta}
\end{center}
\end{figure}

The pupil averaged Mueller matrices obtained from \href{https://roman.ipac.caltech.edu/sims/Coronagraph_inst_param_data_more.html}{IPAC} for wavelength 450-950nm is shown in Figure \ref{Muellermatrix}. We use the Mueller matrix for 575nm to obtain the final Stokes parameters as shown:
\begin{eqnarray}
\left( \begin{array}{c}
1\\ 
q\\ 
u\\ 
v\end{array} \right)\ &=& \left( \begin{array}{cccc} 
M11 & M12 & M13 & M14 \\ 
M21 & M22 & M23 & M24 \\ 
M31 & M32 & M33 & M34 \\ 
M41 & M42 & M43 & M44
\end{array} \right) \left( \begin{array}{c}
1\\ 
q'\\ 
u'\\ 
0\end{array} \right)\ .
\end{eqnarray}
where $q'$ and $u'$ are the stokes parameters obtained after the disk extraction. For measurement of linear polarization ($v'$=0), we consider the 3$\times$3 Mueller matrix terms and estimate $q$, $u$ as: 
\begin{align}
q=M21+M22*q'+M23*u' \\
u=M31+M32*q'+M33*u'
\end{align}
The $p$ and $\theta$ are calculated as,
\begin{align}
p=\sqrt{q^2+u^2} \hspace{0.2cm} \quad 
\theta=0.5\arctan{\frac{u}{q}}.
\end{align}
\begin{align}
\sigma_p=\frac{\sqrt{(q^2\sigma_{q}^2+u^2\sigma_{u}^2})}{q^2+u^2} \hspace{0.3cm} \quad
\sigma_\theta=0.5p\frac{\sqrt{(u^2 \sigma_{q}^2+q^2 \sigma_{u}^2)}}{q^2+u^2}
\end{align}
Figures \ref{stokesqu} and \ref{ptheta} show the final Stokes parameters and polarization fraction along with their errors. We compare the Stokes parameters obtained from MCFOST (Figure \ref{stokes}) for $\epsilon$-Eridani to the final estimated Stokes parameters. The sign of $q$ is found to be flipped due to the Mueller matrix element $M21$. We recover all the Stokes parameters and the polarization fraction after incorporating various error factors in the Roman CGI.

\section{Discussion and future work}
We have presented a mathematical model to simulate polarization simulation of any debris disk through the Roman CGI. As an example, we have modeled the inner ring of the debris disk around Epsilon-Eridani. Various instrument errors, noise, and polarization effects have been accounted for, and we have recovered the input polarization fraction of 0.4$\pm$0.025. The error (2.5\% per pixel) in the polarization fraction measurement is comparable to the expected RMS error for the Roman CGI. Also, this accuracy may suffice for the polarimetry of disks in general, as the linear polarization fraction of disks is in the order of percent to several tens of percent compared to the other stellar objects. However, the achievable measurement accuracy depends on how well the instrument can be calibrated to correct for various polarimetric errors introduced by the instrument optics. Hence a detailed polarization calibration strategy is required for planning the observations of polarized and unpolarized standard stars and performing the required polarization corrections using those observations \cite{millar2016gpi,van2020polarimetric}.  
\par
The \href{https://ipag.osug.fr/~pintec/mcfost/docs/html/overview.html}{MCFOST} model for Epsilon-Eridani is simulated using a single grain composition. We would model the disk by varying dust grain sizes, shapes, and compositions and obtain the wavelength dependence of scattering polarization\cite{tazaki2019effect}. This analysis will enable us to understand how well the disk properties can be constrained using multi-band polarization measurements.

\acknowledgments 
Portions of this work were supported by the WFIRST Science Investigation team prime award \#NNG16PJ24 and the Arizona Board of Regents Technology Research Initiative Fund (TRIF). The authors would like to thank Dr. Max Miller-Blanchaer (UCSB) for the useful discussions and Dr. Bertrand Mennesson (JPL) for generating the Mueller matrices.
 
\bibliography{report} 
\bibliographystyle{spiebib} 

\end{document}